\documentclass[a4paper,12pt]{achemso}
\usepackage[utf8]{inputenc}

\title{Atomistic simulations of the human proteasome inhibited by a covalent ligand}

\author{Michal H. Kol\'a\v{r}}
\affiliation{Max Planck Institute for Biophysical Chemistry, Am Fassberg 11, 37077 G\"ottingen, Germany}
\altaffiliation{University of Chemistry and Technology, Technicka 5, 16628 Prague, Czech Republic}
\email{michal@mhko.science}

\author{Lars V. Bock}
\affiliation{Max Planck Institute for Biophysical Chemistry, Am Fassberg 11, 37077 G\"ottingen, Germany}

\author{Helmut Grubmüller}
\affiliation{Max Planck Institute for Biophysical Chemistry, Am Fassberg 11, 37077 G\"ottingen, Germany}

\usepackage{mathtools}

\begin{document}

\singlespacing

\maketitle

\begin{abstract}
The proteasome is a large biomolecular complex responsible for protein degradation. It is under intense research due to its fundamental role in cellular homeostasis, and tremendous potential for medicinal applications. Recent data from X-ray crystallography and cryo-electron microscopy have suggested that there is a large-scale structural change upon binding of an inhibitor. We carried out atomistic molecular dynamics simulations of the native and inhibited proteasomes to understand the molecular details of the inhibition. Here we describe the technical details of the simulations and assess the quality of the trajectories obtained. The biochemical aspects of the proteasome are under further investigation and will be published elsewhere. This work was a part of the GCS-Prot project at the HLRS, run on the Cray XC40 supercomputing system.
\end{abstract}

\section{Introduction}

Over the many years since the pioneering studies \cite{Levitt75, McCammon77}, biomolecular molecular dynamics (MD) simulations have become a valuable source of scientific data. They capture functional motions of biomolecules with high spatial and temporal resolution and bring information about dynamics and energetics. They complement classic biophysical techniques for structure determination \cite{Trabuco08, Igaev19}, facilitate drug design \cite{Sledz18} or successfully tackle important questions of molecular biology \cite{Lindorff-Larsen11, Kohlhoff14, Kopec18}.

Routinely, microsecond-long trajectories of systems up to few tens of thousands atoms can be achieved on workstations equipped by the customer-class graphical processor units \cite{Kutzner19}. However, simulating larger assemblies and multi-component biomolecular complexes, such as ribosome \cite{Bock18} or proteasome \cite{Wehmer17}, remains a challenge, and can only be done on high-performance supercomputers \cite{Nagel19}, or through distributed computing \cite{Lane13a, Chen19}.

Here, we present results of atomistic MD simulations of human proteasome on a multi-microsecond time scale. The proteasome is a stochastic 2.5 MDa nanomachine responsible for protein degradation in eukaryotic cells via the ubiquitin-proteasome pathway \cite{Saeki12}. It helps maintaining the delicate balance of protein concentrations, and thus plays a fundamental role in cell life cycle. Modulation of proteasome function has a direct effect on cell homeostasis, disruption of which often leads to the cell death \cite{Rechsteiner05}.

\begin{figure}
    \centering
    \includegraphics{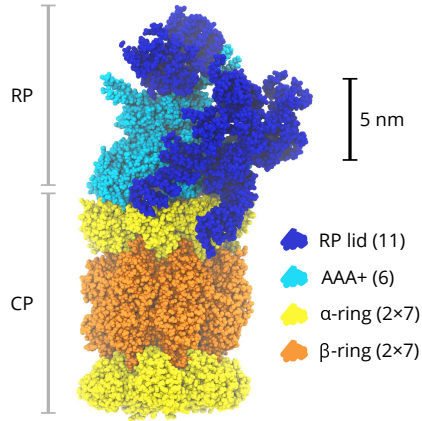}
    \caption{Anatomy of the human proteasome. In the legend, numbers of protein subunits are given in parentheses. Prepared from PDB 5m32 \cite{Haselbach17}.}
    \label{fig:proteasome}
\end{figure}

The proteasome contains two major functional parts: a 20S core particle (CP) and a 19S regulatory particle (RP), which together form the 26S particle depicted in Fig. \ref{fig:proteasome}. The CP is a barrel-shape complex of several protein subunits organized in four rings - two $\alpha$-rings and two $\beta$-rings in a stacked $\alpha\beta\beta\alpha$ arrangement. Three $\beta$-subunits ($\beta_1$, $\beta_2$, and $\beta_5$) have been shown to catalyze the proteolysis. The RP consists of a base and a lid. The base is formed by a ring of six distinct ATPases associated with diverse cellular activities (AAA+), so called regulatory particle triple A proteins. The AAA+ works as an engine that pushes the substrate into the CP \cite{Bar-Nun12}. In addition, several non-ATPase subunits belong to the base and are involved in the recognition of proteasome substrates. Overall, the RP lid consists of eleven different subunits, which recognize and pre-processes protein substrate before it is transferred into CP for degradation.

Large amounts of structural information have been gathered since the proteasome discovery. Recently, several groups have determined the 26S proteasome structure at atomic or near-atomic resolution \cite{Haselbach17,Schweitzer16,Huang16,Chen16}. Despite the continuous efforts, some parts of RP still remain unresolved, mostly due to their high inherent mobility. Understanding the proteasome structure and function poses a fundamental scientific challenge. However, the proteasome is under intensive investigation also due to a tremendous potential for medicinal applications \cite{Bedford11,Kisselev12}.

Oprozomib (OPR) \cite{Demo07} is one of the ligands which have already reached the market as potent anti-cancer agents. The X-ray crystal structure of the CP-OPR complex, resolved at resolution of 1.8 \AA, revealed two OPR molecules in the CP -- one per each of the two $\beta_5$-subunits. Remarkably, the inhibited CP structure is highly similar to the structure of native CP (at 1.9 \AA) \cite{Schrader16}. The protein backbone root-mean-square deviation (RMSD) is only 0.4 \AA, hence the inhibitors do not induce marked structural changes of the CP.

However, based on cryo-electron microscopy (cryo-EM) structures of the 26S proteasome, it has been recently suggested that the RP undergoes a large conformational change upon inhibition \cite{Haselbach17}. When OPR is bound to the CP $\beta$5-subunits, the RP rotates by about 25$^\circ$ into a non-productive state. Intriguingly, the drug binding triggers an allosteric signal which is transferred and amplified over a distance larger than 150 \AA{}, while keeping the (average) structure of the CP almost intact. Our main objective is to understand the atomic details of the CP-RP mutual motion possibly triggered by the OPR binding.

Here we present the computational details of our MD simulations performed on HLRS Hazel Hen, and assess the quality of the trajectories obtained.

\section{Methods}

\subsection{Simulated Systems}

\begin{figure}
    \centering
    \includegraphics{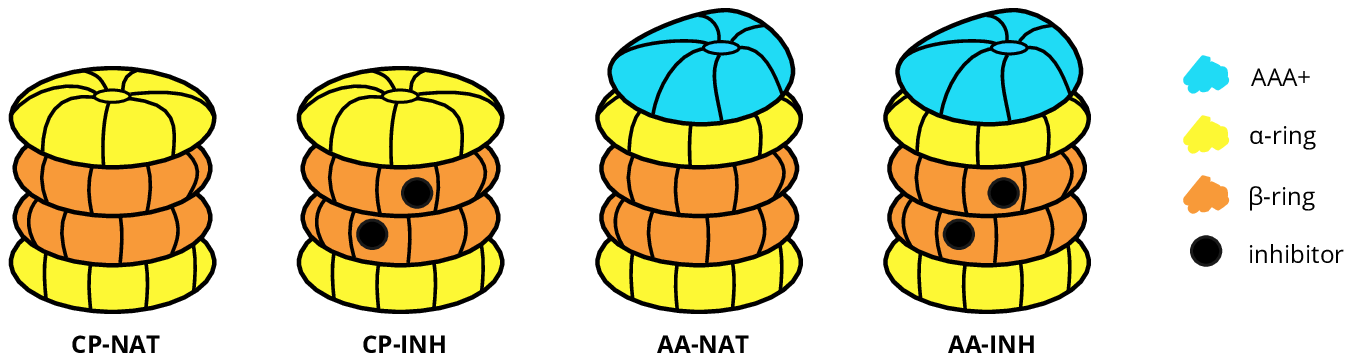}
    \caption{A scheme of the four proteasome constructs simulated.}
    \label{fig:constructs}
\end{figure}

We have built four proteasome constructs: native CP, inhibited CP, native AA, and inhibited AA, where AA stands for a CP with one AAA+ ring (Fig. \ref{fig:constructs}). The inhibited AA structure was prepared from the available cryo-EM data (PDB 5m32 \cite{Haselbach17}). The dangling N-terminal $\alpha$-helices of the AAA+ regulatory subunits 6A, 6B, 8 and 10 (UniProt naming convention) were omitted. Where needed, the CP subunits were completed by missing amino acids to keep the up-down sequence symmetry of the pairs of $\alpha$- and $\beta$-rings. By removing the two inhibitors from the inhibited AA, we prepared the native AA structure. Experimentally determined coordinates of water molecules, K$^+$, Mg$^{2+}$ and Cl$^-$ ions were taken from the X-ray data (PDB 5le5 \cite{Schrader16}) and added after superimposing the backbone atoms of the CP. Each of the six AAA+ subunits contained one adenosine diphosphate as found in the cryo-EM model.

Each construct was placed into a periodic rhombic dodecahedron box of sufficient size such that the distance between the solute and box faces was not shorter than 1.5 nm. The system was dissolved in a solution of K$^+$ and Na$^+$ ions of the excess concentrations of 139~mM and 12~mM, respectively, and neutralized by Cl$^-$ anions. In total, the simulations contained about 0.8 and 1.6 million atoms, for the CP and AA constructs, respectively.

\subsection{Simulation Setup}

Molecular dynamics simulations were carried out using classical interaction potentials. A standard version of the Amber force field was used for the proteasome \cite{Lindorff-Larsen10}. The OPR was parametrized together with its covalently bound N-terminal threonine in the spirit of the Amber family of force fields, using fitted partial atomic charges from the Restricted Electrostatic Potential method \cite{Bayly93}, and General Amber Force Field parameters \cite{Wang04a}. The TIP3P water model \cite{Jorgensen83} and ion parameters by Joung and Cheatham \cite{Joung08} were used.

Newton's equations of motion were integrated using the leap-frog algorithm. All bonds were constrained to their equilibrium lengths using the parallel LINCS algorithm of the sixth order \cite{Hess08}. Hydrogen atoms were converted into virtual sites \cite{Berendsen84} which allowed using 4-fs integration time step in the production simulations. Electrostatic interactions were treated by the Particle Mesh Ewald method \cite{Darden93} with the direct space cut-off of 1.0~nm and 0.12~nm grid spacing. Van der Waals interactions, described by the Lennard-Jones potential, used a cut-off of 1.0~nm.

The systems were equilibrated in several steps. First, each system was thoroughly energy minimized. In some instances, the virtual-site model caused crashes, thus for the minimization a model with explicit hydrogen atoms and flexible water molecules was used. Second after minimizing water molecules in roughly 50,000 steps of the steepest descent algorithm, the water was heated from 10 K to 300 K in a 5-ns long constant-volume MD simulation, where two thermostats were used separately for the solute and solvent. Velocities were selected randomly from the Maxwell-Boltzmann distribution at the given temperature. Moreover during heating, the solute heavy atoms were restrained by harmonic potential to their starting coordinates with the force constant of 5000 kJ\,mol$^{-1}$\,nm$^{-2}$. Next, the density of the system was equilibrated in a 20-ns long constant-pressure MD simulation at 300~K and 1~bar, where v-rescale thermostat \cite{Bussi07} and Berendsen barostat \cite{Berendsen84a} were used. During this step, the solute was still restrained. Finally, the position restraints were gradually released in a 50 ns-long simulation, where the force constant was interpolated between its initial value and zero.

In production runs, the isobaric-isothermal statistical ensembles were generated at 300 K and 1 bar using the v-rescale thermostat and Parrinello-Rahman barostat \cite{Parrinello81}, respectively. We simulated roughly 4 microseconds per trajectory with the exception of the CP-INH, where the length was increased up to 5.6 microseconds. For validation of the observed phenomena, another set of simulations was carried out. This started from the final conformation of the inhibited proteasome simulations, where the inhibitor was removed. Due to technical reasons, only the coordinates of the solute were kept, whereas the water and ions were added from scratch in the same manner as with the simulations initiated from the experimental conformation.

\subsection{Software Details}

The simulations were carried out in the GROMACS 2016 package \cite{Abraham15}. It is a well-established, highly-optimized C/C++ code released under Lesser General Public License. Initially, we used the standard module available on HLRS Hazel Hen supercomputer. After removal of version 2016 from the list of supported modules, we used a self-compiled version with very similar performance characteristics.

\begin{figure}
    \centering
    \includegraphics{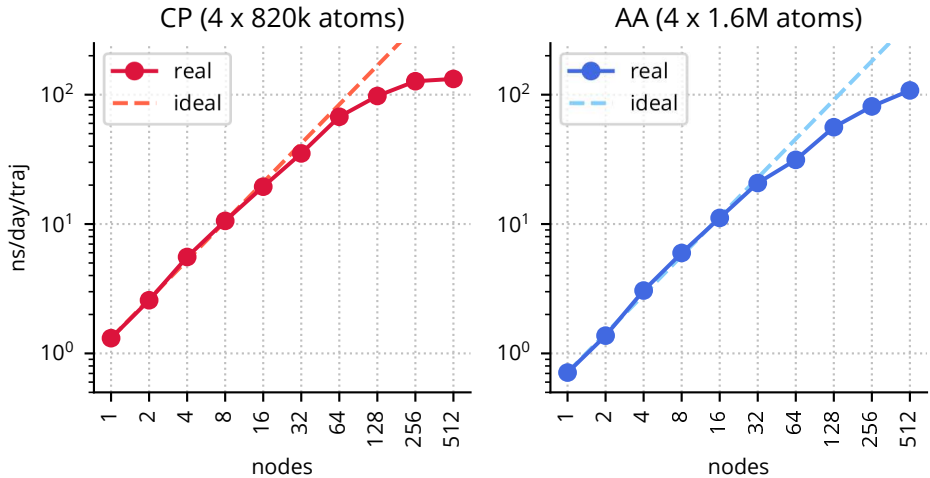}
    \caption{Performance of GROMACS 2016 on the Hazel Hen supercomputer.}
    \label{fig:scaling}
\end{figure}

GROMACS uses a mixed MPI/OpenMP parallelization which may scale down to ``few tens of atoms per core'' \cite{Abraham15}. In our case, the scaling was better for the larger AA than smaller CP construct (Fig. \ref{fig:scaling}). For the production runs, we employed 128 or 256 nodes with two 12-core Intel Xeon (Haswell gen.) processors each. For each system, four independent trajectories initiated with different velocities from the Maxwell-Boltzmann distribution were generated. To improve scaling, bundles of the four simulations were run as a single \emph{aprun} argument.

Memory requirements were rather low, our system consumed about 600 MB of memory per MPI task (for the AA construct) comprising several OpenMP threads.

The simulations generated large amounts of data compressed to a high degree at the level of GROMACS output routines. GROMACS allows checkpoint-file dependent restarts of the simulations so the simulations were run as chained jobs to fit the maximum runtime of 24 hours on Hazel Hen. Due to our interest in the solute behavior and limited disk space, we saved the water coordinates less frequently (100 ps) than the coordinates of the solute (10 ps). Each chain step produced a portion of solute trajectory of about 5.5 GB. Due to the limited disk space, these portions were downloaded frequently to our local servers, and concatenated before the final analysis.

\section{Results and Discussion}

For all trajectories, we calculated the root-mean-square deviations (RMSDs) of the backbone atoms with respect to the starting proteasome conformation according to Eq.~\ref{eq:rmsd}. The analysis was performed after a least-square alignment of the trajectory to the starting conformation using the backbone atoms of the CP subunits.

\begin{equation}
\mathrm{RMSD}(t) = \sqrt{\frac{1}{N_a} \sum_a^{N_a}
\left ( \mathbf{r}_a(t) - \mathbf{r}_a(0) \right )^2},
\label{eq:rmsd}
\end{equation}
where $N_a$ is the number of atoms in a trajectory, the $\mathbf{r}_a(0)$ is position vector at time 0, i.e.  the experimental structure, and $\mathbf{r}_a(t)$ is the position vector at time $t$.

\begin{figure}
    \centering
    \includegraphics{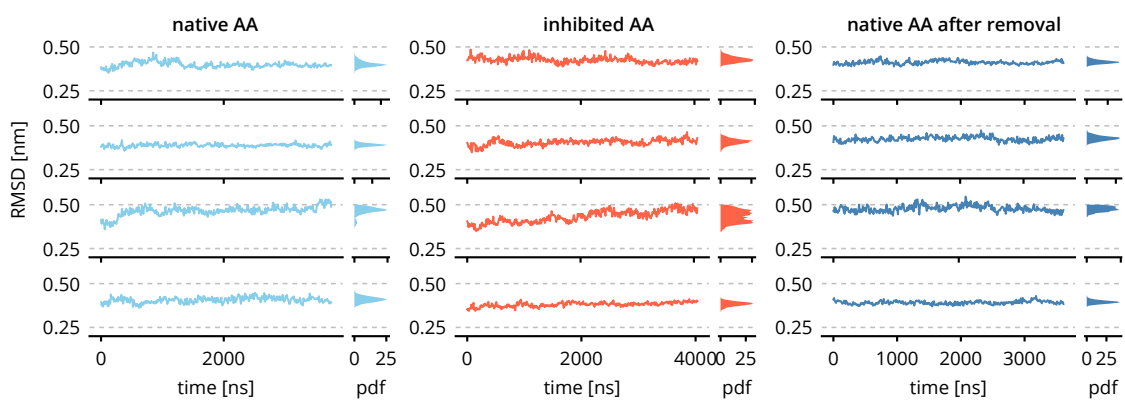}
    \caption{Time evolution of the backbone root-mean-square deviation (RMSD) and the respective probability density functions (pdf) obtained for the AA constructs. Four independent trajectories of the inhibited constructs are shown in red, the native in blue. The pale blue traces started from the experimental conformation, whereas the dark blue started from the final conformations of the inhibited constructs.}
    \label{fig:rmsdTimeAA}
\end{figure}

All of the trajectories appear stable within the limits of such a simple measure as RMSD. Fig.~\ref{fig:rmsdTimeAA} shows the RMSD profiles with the averages over respective trajectories between 0.30 and 0.35 nm. These values are expected, given the size of the system (over 6,000 amino acids) and no significant drift. Similar plots for AA constructs are in Fig.~\ref{fig:rmsdTimeAA}. Here the RMSD values averaged over the trajectories are around 0.40 nm, with two instances higher than 0.45 nm. The profiles show no significant drift. Higher RMSD values are related to the size of the system (over 8300 amino acids), and to the fact that only the CP subunits were used for the alignment.

\begin{figure}
    \centering
    \includegraphics{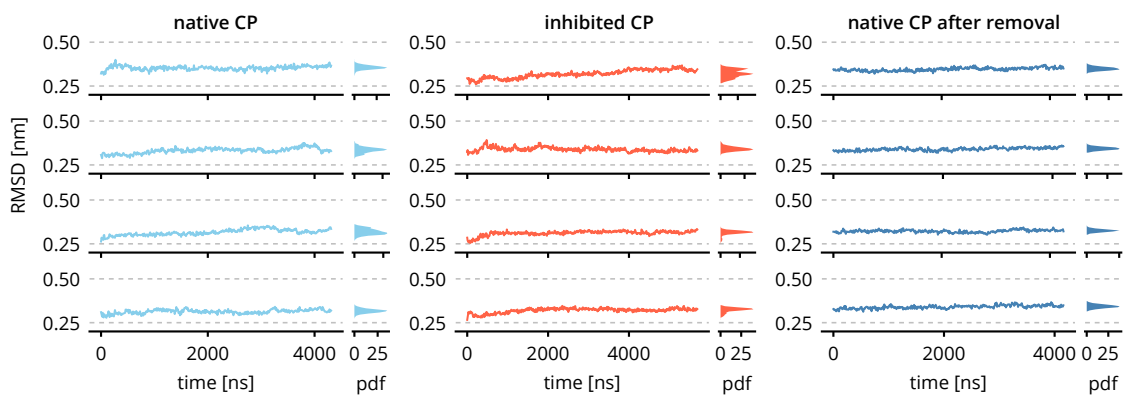}
    \caption{Time evolution of the backbone root-mean-square deviation (RMSD) and the respective probability density functions (pdf) obtained for the CP constructs. Four independent trajectories of the inhibited constructs are shown in red, the native in blue. The pale blue traces started from the experimental conformation, whereas the dark blue started from the final conformations of the inhibited constructs.}
    \label{fig:rmsdTimeCP}
\end{figure}

The natural sequence symmetry of the CP allowed us to assess the convergence of the simulations. The CP is free to move in the simulation box, so the structure and dynamics of the subunits in the upper and lower halves should converge to common values if the free-energy minimum is well defined.

For the CP constructs, we calculated the average conformation between 1600 and 3600 ns of each trajectory, i.e. the mean position vector x of all heavy atoms. We aligned the upper and lower halves using the backbone atoms. Then for each pair of equivalent subunits in the upper and lower half, the $\mathrm{RMSD_{u-l}}$ was calculated as follows.

\begin{equation}
\mathrm{RMSD_{u-l}} = \sqrt{\frac{1}{N_a} \sum_{a}^{N_a}
\left ( \mathbf{r}_{a,u} - \mathbf{r}_{a,l} \right )^2},
\end{equation}
where $\mathbf{r}_{a,u}$ is the position vector of atom $a$ in the upper half, the $\mathbf{r}_{a,l}$ is the position vector of the equivalent of atom $a$ in the lower half, and the sum runs over $N_a$ atoms on one proteasome half. If the mean structures were identical as proposed by the sequence symmetry, the $\mathrm{RMSD_{u-l}}$ would be zero. Non-zero values indicate structural variation between equivalent subunits.

\begin{figure}[t]
    \centering
    \includegraphics{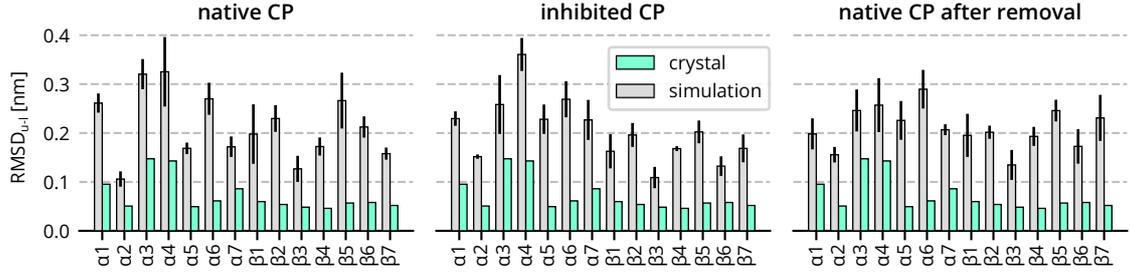}
    \caption{The $\mathrm{RMSD_{u-l}}$ of the CP constructs. Values from the crystal structure are shown in green. The gray bars represent error bars represent standard errors of the mean obtained from four independent trajectories.}
    \label{fig:rmsdChains}
\end{figure}

Fig. \ref{fig:rmsdChains} shows the $\mathrm{RMSD_{u-l}}$ of CP constructs obtained from the experimental structure and from the simulations. The non-zero values in the experimental structures may be related to crystal-packing effects. Moreover, there are number of surface protein loops and terminals which are flexible. Thus, the free-energy surface is expected to feature many shallow minima, so their conformation in the upper and lower halves may vary. The $\mathrm{RMSD_{u-l}}$ values from the simulations are higher, for $\beta$-units by factor of about 3, for $\alpha$-units by factor of about 2.

\begin{figure}
    \centering
    \includegraphics{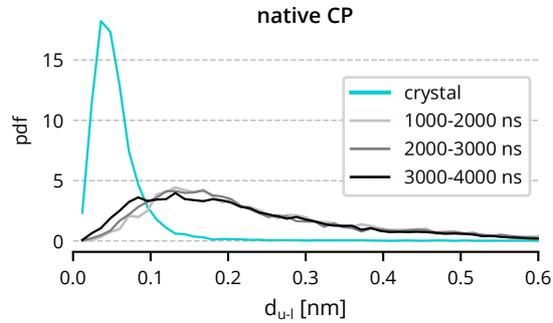}
    \caption{Histogram of distances between equivalent backbone atoms of the upper and lower halves in the crystal structure (green) and simulation (gray). Data from various parts of one trajectories are shown in shades of gray.}
    \label{fig:distHisto}
\end{figure}

\begin{figure}
    \centering
    \includegraphics{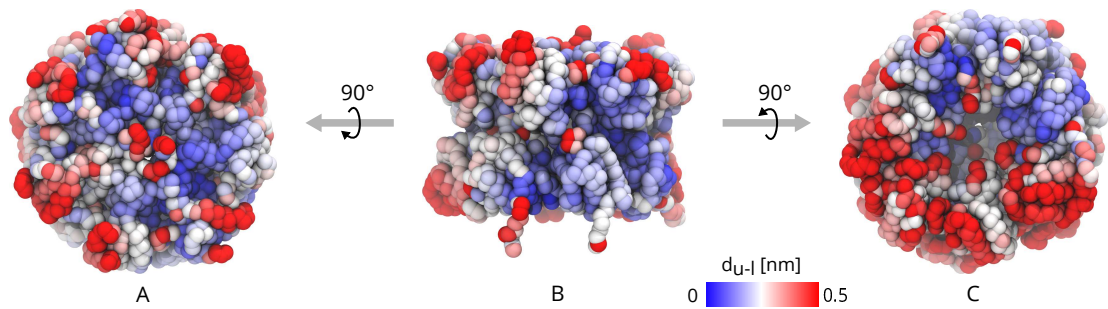}
    \caption{Projection of the distance $d_{u-l}$ between equivalent atoms of the upper and lower halves onto the crystal structure of the proteasome oriented towards the $\alpha$-ring (A), front face (B) and $\beta$-ring (C). Each sphere represents one amino acid and the color scale goes from blue (low $d_{u-l}$) through white to red (high $d_{u-l}$). }
    \label{fig:distProj}
\end{figure}

Further after the least-square alignment of the backbone atoms of the two proteasome halves, we calculated distances $d_{u-l}$ between equivalent atoms in upper and lower halves. The histogram of $d_{u-l}$ shows (Fig. \ref{fig:distHisto}) a maximum about 0.05 nm for the crystal. For simulation, the maximum lies slightly beyond 0.1 nm and is broader. This indicates that the proteasome conformations averaged over a trajectory are structurally less symmetric than the crystal. A projection of the simulation $d_{u-l}$ onto the proteasome structure (Fig. \ref{fig:distProj}) reveals that the largest structural variations are located in the surface loops and terminal chains. In the course of simulation time, the $d_{u-l}$ profiles do not diverge (Fig. \ref{fig:distProj}), or even slightly improve towards shorter values.

\section{Concluding Remarks}

Using the Cray XC40 supercomputer, we have performed atomistic MD simulations of the proteasome, a multi-protein complex responsible for protein degradation recently used as an anti-cancer drug target. We obtained trajectories totalling 100 $\mu$s in length of several systems with 0.8 and 1.6 million atoms.

Here, we have presented a technical report focused on the simulation setup, run-time performance and basic analyses. Next, we will focus on proteasome function and its regulation and will present such biochemical aspects in future texts. The trajectories obtained through the MD simulations are likely of sufficient quality to explain at atomic level what changes the inhibitor OPR triggers and how these can modulate proteasome function.

\bibliography{refs}

\end{document}